\documentclass[a4paper,twocolumn,11pt,accepted=2022-04-12]{quantumarticle}

\pdfoutput=1

\usepackage[utf8]{inputenc}
\usepackage[english]{babel}
\usepackage[T1]{fontenc}
\usepackage{amsmath}
\usepackage{amsthm}
\usepackage{amssymb}
\usepackage{hyperref}
\usepackage{braket}

\usepackage{tikz}
\usepackage{lipsum}
\usepackage{epstopdf}

\usepackage{empheq}
\usepackage{bbm}

\newcommand{\tr}{\textrm{Tr}}

\graphicspath{ {images/}{images/figure1/}{images/figure2/}{images/figure3/}{images/figure4/}{images/figure5/} }

\newtheorem{innercustomgeneric}{\customgenericname}
\providecommand{\customgenericname}{}
\newcommand{\newcustomtheorem}[2]{%
	\newenvironment{#1}[1]
	{%
		\renewcommand\customgenericname{#2}%
		\renewcommand\theinnercustomgeneric{##1}%
		\innercustomgeneric
	}
	{\endinnercustomgeneric}
}

\newcustomtheorem{customdef}{Definition}

\newcommand{\ii}{\mathrm{i}}

\begin{document}

\title{Eigenstate entanglement in integrable collective spin models}

\author{Meenu Kumari}
\email{mkumari@uwaterloo.ca }
\affiliation{Perimeter Institute for Theoretical Physics, Waterloo, ON N2L 2Y5, Canada}

\author{\'Alvaro M. Alhambra}
\email{alvaro.m.alhambra@gmail.com}
 \affiliation{Max-Planck-Institut fur Quantenoptik, D-85748 Garching, Germany}

\begin{abstract}
The average entanglement entropy (EE) of the energy eigenstates in non-vanishing partitions has been recently proposed as a diagnostic of integrability in quantum many-body systems. For it to be a faithful characterization of quantum integrability, it should distinguish quantum systems with a well-defined classical limit in the same way as the unequivocal classical integrability criteria. We examine the proposed diagnostic in the class of collective spin models characterized by permutation symmetry in the spins. The well-known Lipkin-Meshov-Glick (LMG) model is a paradigmatic integrable system in this class with a well-defined classical limit. Thus, this model is an excellent testbed for examining quantum integrability diagnostics. First, we calculate analytically the average EE of the Dicke basis $\{|j,m\rangle \}_{m=-j}^j$ in any non-vanishing bipartition, and show that in the thermodynamic limit, it converges to $1/2$ of the maximal EE in the corresponding bipartition. Using finite-size scaling, we numerically demonstrate that the aforementioned average EE in the thermodynamic limit is universal for all parameter values of the LMG model. Our analysis illustrates how a value of the average EE far away from the maximal in the thermodynamic limit could be a signature of integrability.
\end{abstract}

\maketitle

\section{Introduction}

While the notion of integrability in classical mechanics is well understood through the connection between degrees of freedom and constants of motion (in the sense of Liouville integrability) \cite{arnol2013mathematical,babelon_bernard_talon_2003}, a fully consistent and rigorous notion remains elusive for quantum systems \cite{caux2011remarks}. Quantum integrability is usually associated with the existence of an exact solution of the model, for instance based on Yang-Baxter equation \cite{Baxter1982,gaudin_2014}, such as the Bethe ansatz \cite{bethe1931theorie}, or with other features, such as a set of simple conserved quantities, or Poissonian level statistics \cite{berry1977level}. However, none of these features provide an unambiguous characterization of integrable quantum systems and distinguish them from nonintegrable quantum systems. Since the quintessential goal is to characterize quantum dynamics, a good criterion or measure should be able to unequivocally split all quantum models into two distinct classes - integrable and nonintegrable - each with fundamentally distinct dynamical behaviour \cite{caux2011remarks}.

Quantum entanglement has the potential of constituting a defining measure that can characterize integrability in the class of quantum systems with an underlying tensor product structure. This is supported by numerous studies that has directly linked entanglement with dynamical features \cite{lewis2019dynamics}. A prominent example is the qualitative behavior of the spread of entanglement in different types of spin systems.  This includes integrable and non-integrable chains \cite{Chiara_2006,cincio2007entropy,fagotti2007evolution,kim2013ballistic,castro2020entanglement}, many-body localized systems \cite{bardarson2012unbounded,serbyn2013universal,Friesdorf_2015} or quantum scars \cite{turner2018quantum,ho2019periodic}, or other models such as the Dicke model \cite{Lewis_Swan_2019} and the quantum kicked top \cite{chaudhury2009quantum,Neill_2016}. Another possibility to explore this connection is through the study of entanglement in the energy eigenstates, and how this directly affects quantum dynamics. 
A succinct way of studying entanglement entropy (EE) of all the energy eigenstates in bipartite systems is to look at their uniform \emph{average} 
\begin{equation}
\bar S_A \equiv \frac{1}{d} \sum_{k=1}^d S(\text{Tr}_{B}{\ket{E_k}\bra{E_k}}),
\end{equation}
where $S(\rho)=-\tr{\rho \log \rho}$ refers to the von Neumann entropy \cite{nielsen2010quantum} which is a measure of entanglement in bipartite pure states. The sum is over all the eigenstates of the total Hamiltonian $H_{AB}=\sum_k E_k \ket{E_k}\bra{E_k}$. 

The average EE has been recently studied in a number of lattice models in the context of eigenstate thermalization hypothesis (ETH) \cite{beugeling2015global,vidmar2017entanglement2,murthy2019structure,huang2019universal}. These works have verified that the EE of the eigenstates (and hence their average) in chaotic systems is close to the maximum possible EE, such that, for a bipartite system of dimension $d_A \times d_B$ with $d_A \le d_B$, $\bar S_A \simeq S_{\max} \simeq \log d_A $ in the thermodynamic limit. This suggests that the eigenstates of these models resemble random states, for which their EE is near maximal on average $\bar S_A / S_{\max} \rightarrow 1$ \cite{page1993average}. 
 
This average has been shown to be very different in a few known integrable models in one dimension. At half bipartition $p\equiv \frac{N_A}{N_A + N_B}=\frac{1}{2}$ (where $N_A$ and $N_B$ are the number of qubits in the respective subsystem), it has been analytically shown that
$0.52 < \frac{\overline{S}_A}{S_{\max}} <0.59$ for translation-invariant free fermions \cite{vidmar2017entanglement}, for the XY chain \cite{hackl2019average} and in \cite{lydzba2020eigenstate} to be $2-\frac{1}{\ln(2)}\sim 0.557$ for random quadratic (integrable) models. Perhaps more surprisingly, the finite-size scaling analysis in \cite{leblond2019entanglement} showed that the average EE of the \emph{interacting} integrable XXZ model converges to the free-fermionic value in the thermodynamic limit.

Importantly, this departure from maximal arises for bipartitions proportional to the system size $N_A \propto N_A + N_B $. On the contrary, the average EE for small bipartitions $ N_A \ll N_B$ has been shown to coincide for both integrable and nonintegrable lattice systems \cite{keating2015spectra,huang2019universal,leblond2019entanglement} (with the exception of localized ones \cite{Bauer_2013,Friesdorf_2015}). Studies of connections between chaos or integrability and entanglement in small (vanishing) bipartitions have led to conflicting conclusions \cite{lombardi2011entanglement,Vaibhav2015,lombardi2015,ruebeck2017entanglement,kumari2019untangling,dogra2019quantum} demonstrating the importance of the choice of the bipartition.

The aforementioned studies suggest that a fixed average EE far from the maximal for large (non-vanishing) bipartitions may be determined by the integrability of the model. Here, we study this conjecture in integrable systems of collective spin models, whose Hilbert space is the symmetric subspace of $N$-spins \cite{harrow2013church, stockton2003characterizing}. In particular, we focus on the paradigmatic integrable Lipkin-Meshkov-Glick (LMG) model \cite{lipkin1965,meshkov1965,glick1965,Enrique2006}. We show that the half-bipartition average EE in LMG converges to a universal value, which we calculate analytically for specific parameters in the LMG Hamiltonian as the average EE over the Dicke basis. We find that the average value of the entanglement entropy greatly diverges from the maximal, as is the case for other previously studied integrable models \cite{vidmar2017entanglement,hackl2019average,lydzba2020eigenstate,leblond2019entanglement}. We expect that the LMG model captures the average entanglement behaviour of a larger class of collective spin models, as is the case for their ground states \cite{vidal2007entanglement}, as well as nonlinear LMG models \cite{Deutsch2021,Poggi2020}. 

Collective spin models have a well-defined classical limit where they can be categorized as integrable or nonintegrable as per the unequivocal integrability criteria of classical physics. This makes them excellent testbeds for the diagnostics related to quantum integrability. Being a collective spin model, the LMG model also has such a well-defined classical limit in which it is integrable \cite{Santos2018,Chinni2021}. This is a relevant difference with respect to previous results based on lattice models \cite{vidmar2017entanglement,leblond2019entanglement,lydzba2020eigenstate}. Additionally, the LMG is also known to be quantum-integrable using Bethe ansatz \cite{Richardson1963277,Richardson1964221} which is one of the main integrability definitions used in quantum mechanics.

Moreover, we analytically compute the thermodynamic limit of the average EE in the Dicke basis for any \emph{non-vanishing} bipartition $p>0$, in addition to $p=1/2$. We show that it converges to half the value of maximal EE in the corresponding bipartition irrespective of the value of $p$ when it is nonzero. We also show  numerically that this holds in the LMG model. This is in stark contrast to other integrable lattice models previously studied \cite{hackl2019average,lydzba2101entanglement,lydzba2020eigenstate,leblond2019entanglement}. For these, the coefficient of the leading term in volume law of entanglement has been shown to be dependent on the fraction of the bipartition. 

We also go beyond the average EE and study the entanglement distribution over the eigenstates. We find that it displays a variety of structures depending on the specific parameters, with singular points which correspond to singularities in the density of states \cite{ribeiro2007thermodynamical,ribeiro2008exact}. Given these very different EE distributions, it is noteworthy that the average EE always converges to the same value.

The paper is structured as follows. First, in Sec. \ref{sec:Dicke}, we calculate the average EE of the Dicke basis, which is the eigenbasis of the LMG model for certain parameter sets. In Sec. \ref{sec:average} we introduce the model and show the numerical calculations of the average EE. In Sec. \ref{sec:distribution} we study the distributions of EE as a function of the energy, and then conclude in Sec. \ref{sec:conclusion}.


\section{Average entanglement entropy of Dicke basis}\label{sec:Dicke}

\begin{figure}
\centering\includegraphics[width=0.45\textwidth]{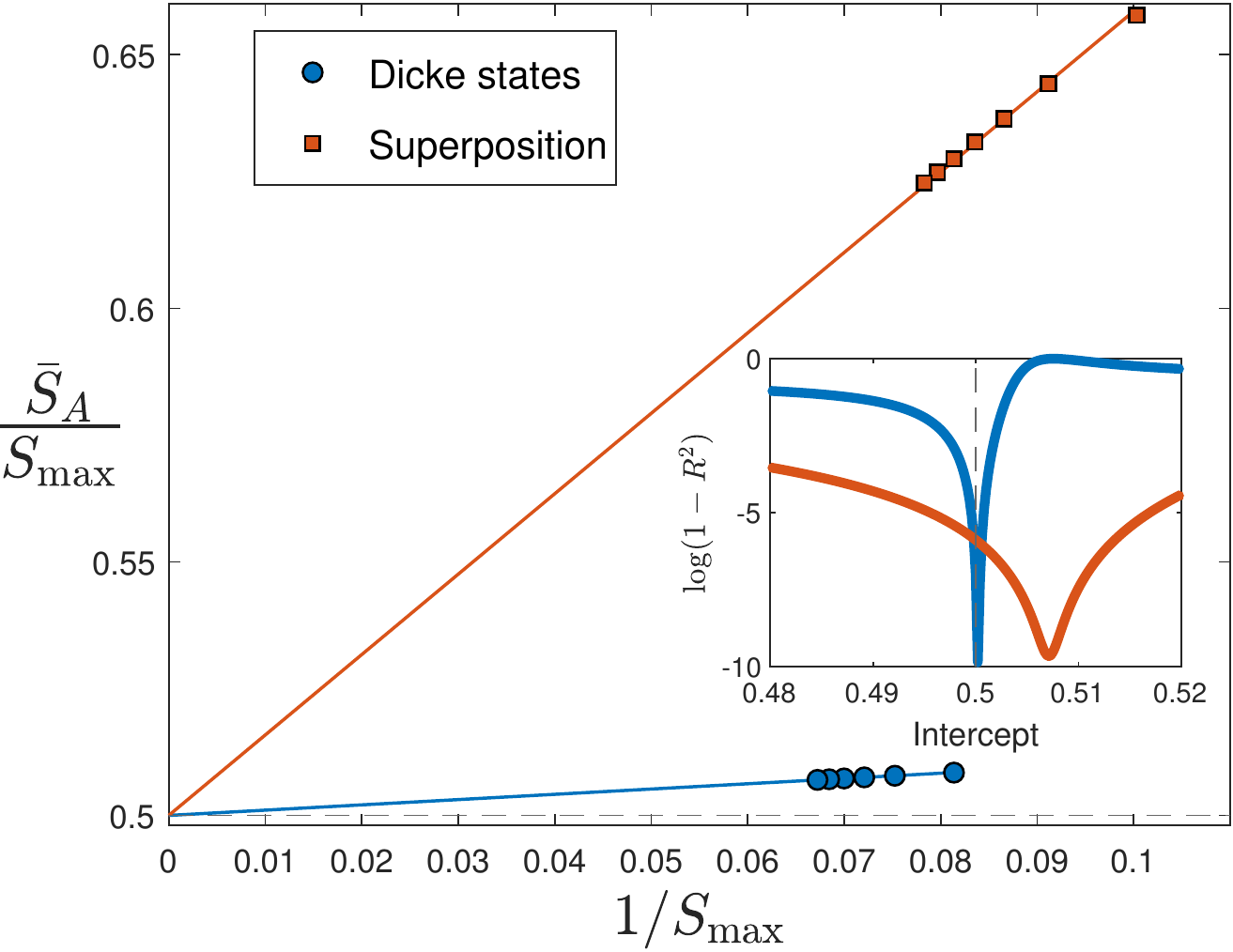}
\caption{Normalized average entanglement for the complete Dicke basis and the basis consisting of equal superposition of conjugate Dicke states as a function of $1/S_{\text{max}} (\equiv 1/\log(N/2+1))$ at half bipartition. The linear fits correspond to $a+b/S_{\text{max}}$ with intercept $a$ fixed to $1/2$. $N \in [10^4,6\times 10^4]$ for the Dicke basis, and $[4\times 10^3, 2.8\times 10^4]$ for the superposition basis for the shown data points. Inset shows $\log (1-R^2)$, where $R^2$ is the coefficient of determination of the linear fit, for different fixed values of the intercept $a$. At $a=1/2$, the values of $(1-R^2)$ are $10^{-10}$ and $10^{-5}$ for the Dicke basis and the superposition basis, respectively.}
\label{fig:EEDicke}
\end{figure}

Let us consider a system of $N{=}2j$ spin-$1/2$ qubits whose Hilbert space is $2^N$ dimensional. The subspace corresponding to the maximum total spin $j$ is an $N+1$ dimensional ``symmetric" subspace (under permutation of the qubits). It is spanned by the Dicke basis $\{|j,m\rangle \}_{m=-j}^j$ ($ \equiv \{|N,k\rangle \}_{k=0}^N$) which can be written in the computational basis as
\begin{equation}
\ket{j,m}=\binom{2j}{j-m}^{-1/2}\sum_{l=1}^{\binom{2j}{j-m}} \hat P_l \ket{\underbrace{0..0}_{j+m} \underbrace{1..1}_{j-m}},
\end{equation}
where $\{\hat P_l\}$ represent the $\binom{2j}{j-m}$ non-trivial permutations, and $\left(j-m\right)$ represents the number of 1's in any term in the $\ket{j,m}$ state \cite{Dicke1954,moreno2018all}. In terms of the collective spin operators, the Dicke basis is the simultaneous eigenbasis of the total angular momentum squared operator, $J^2 = J_x^2+J_y^2+J_z^2$, and $J_z$. These operators are given by
\begin{align} \label{SpinOperators}
    J_i = \frac{1}{2}\sum_{k=1}^N \sigma_k^i, \hspace*{4mm} i \in \{x,y,z\},
\end{align} 
where $\sigma^i$'s are the single qubit Pauli matrices, and obey the commutation relation $[J_j,J_j] = \ii \epsilon_{ijk}J_k$.

For any bipartition $N_A{:}N_B$ of $N$ qubits in a Dicke state, there exists a simple closed-form expression of the Schmidt coefficients, in terms of the hypergeometric distribution \cite{latorre2005entanglement,moreno2018all}
\begin{equation}
\lambda_q= \frac{{{N_A}\choose{q}} {{ N_B}\choose{j-m-q}}} {  {{2j}\choose{j-m}}}, \quad q \in{0,1,...,j-m}.
\label{SchmidtCoeffDicke}
\end{equation}
This allows us to analytically compute their entanglement entropy using approximations to the entropy of the hypergeometric distribution \cite{cichon2013delta}, 
\begin{align}
S_A(\ket{j,m})&=-\sum_q \lambda_q \log \lambda_q \\&\simeq
\frac{1}{2}\log(4\pi e j p_1^* p_2^*),\label{eq:approxhyp}
\end{align}
where $p_1^*=\frac{j^2-m^2}{4j^2}$ and $p_2^*=\frac{N_A  N_B }{4j^2}$, and $e$ is Euler's number. In all our analytical and numerical calculations, $\log$ implies $\log$ base 2. We use this approximation to obtain a bound on the average EE over all Dicke states
\begin{equation}
    \bar{S}_A=\frac{1}{2j+1}\sum_{m=-j}^j S_A(\ket{j,m}).
\end{equation} 

Without loss of generality, we assume $N_A\leq N_B$. For any non-vanishing finite fraction $p=\frac{N_A}{N}\le \frac{1}{2}$ of the system, the average entropy in subsystem $A$ is upper bounded by
\begin{equation}\label{eq:entupper}
     \overline{S}_A \le \frac{1}{2} \log \left(\pi e j p(1-p) \right)+ \mathcal{O}\left( j^{-1}\right).
\end{equation}
This follows from the fact that the Dicke state with $m=0$ is the one with the largest EE, $S_A \sim \frac{1}{2}\log (\pi e j p(1-p))$, and thus upper bounds all the other terms in the sum.

On the other hand, the average EE is lower bounded by
\begin{equation}\label{eq:entlower}
    \overline{S}_A \ge \frac{j}{(2j+1)}  \log \left(\frac{\pi}{2e} j p(1-p) \right) + \mathcal{O}\left( j^{-1/2}\right).
\end{equation}
To prove this, we approximate the EE of most eigenstates with Eq.~\eqref{eq:approxhyp}, and then evaluate the average with the Euler-McLaurin formula. This allows us to approximate the sum over integers $\{m\}$ with a simple integral in a way that the error is controlled and decreasing with $j$. The lower bound comes from the fact that this approximation does not apply to Dicke states with $\vert m \vert \sim j$, which we simply omit in the sum over $m$. We see that both bounds Eq. \eqref{eq:entupper} and \eqref{eq:entlower} match up to subleading terms in the limit $j \rightarrow \infty$. 
The detailed proofs can be found in Appendix \ref{app:dickehalf}.

These values of the EE are far from maximal. Subsystems of permutation-symmetric multi-qubit systems are also permutation-symmetric, and hence the dimension of their local Hilbert space $A$ is $N_A +1$ (as opposed to $2^{N_A}$). This means that the maximum possible EE in the bipartition $A{:}B$ is $S_{\max} = \log( N_A +1) =\log( pN+1)$ (given $N_A \leq N_B$) \cite{stockton2003characterizing}. From this, we see that the average EE over Dicke states converges to exactly half of the maximum possible EE in the thermodynamic limit, as
\begin{equation}\label{eq:entone}
    \lim_{j \rightarrow \infty } \frac{\overline{S}_A}{S_{\max}}=\lim_{j \rightarrow \infty } \left(\frac{1}{2}+\mathcal{O}\left(\frac{1}{S_{\max}}\right) \right)=\frac{1}{2}.
\end{equation}
Moreover, this value is independent of $p$. Henceforth, we will refer to $\frac{\overline{S}_A}{S_{\max}}$ as `normalized average EE', where the normalization is provided by maximum possible EE in the corresponding bipartition $S_{\max}=\log(pN+1)$. Notice that this type of entanglement scaling is much slower than what is found in generic states in the whole Hilbert space, for which the entanglement entropy can take values up to $S_A= N_A \log 2$.

We now also study the numerical behavior of the convergence of normalized average EE over Dicke basis in the thermodynamic limit. We carry out a numerical finite-size scaling analysis, and show the results for $p=1/2$ in Fig. \ref{fig:EEDicke} where we see the numerical convergence to $1/2$ (with negligible finite-size effects as measured by the coefficient of determination, as explained in the next section). We have observed the same to be numerically true for the quarter bipartition $p=1/4$ as well.

In addition to the Dicke basis $\{|j,m\rangle \}_{m=-j}^j$, we carry out the finite-size scaling analysis of the following basis formed out of equal superposition of conjugate Dicke states, that is, $\{ \{ \frac{1}{\sqrt{2}} (|j,m\rangle \pm |j,-m\rangle) \}_{m=1}^j, \ket{j,0} \}$. Such a superposition basis (up to a global rotation) as well as the Dicke basis are eigenbases of the LMG model for special sets of parameter values as explained in the next section. For $p=1/2$ and $\vert m \vert> \frac{j}{2}$, the reduced state of subsystem A, $\rho_A$, can be diagonalized into two blocks with the same eigenvalues, and hence the half-bipartition EE can be computed analytically as
\begin{equation}\label{eq:SupDicke}
S_A(\frac{1}{\sqrt{2}} (|j,m\rangle \pm |j,-m\rangle)) = S_A(|j,m\rangle) + 1.
\end{equation}
For $p=1/2$ and $m \leq \frac{j}{2}$, $\rho_A$ cannot be diagonalized analytically in such simple form, and hence analytical EE calculation seems intractable. The finite-size scaling analysis of normalized average EE for the equal superposition Dicke basis for $p=1/2$ and $p=1/4$ yields convergence to $1/2$ in the thermodynamic limit. The $p=1/2$ case is shown in Fig. \ref{fig:EEDicke}.

Besides non-vanishing bipartitions, we also study the average EE in a special case of vanishing bipartition, when subsystem $A$ is a single qubit. For the Dicke basis, we obtain $ \substack{\lim \\ {j \rightarrow \infty}} \frac{\overline{S}_A}{S_{\max}} = \frac{\log{e}}{2} \simeq 0.7213 $ (derived in Appendix \ref{app:1QubitDicke}). On the other hand, for the equal superposition Dicke basis, the 1-qubit average EE is always 1 (Appendix \ref{app:1QubitDicke}). Since the thermodynamic limits of the 1-qubit average EE do not coincide for two different bases, both of which are eigenbasis of the integrable LMG model for specific choice of parameters, the possibility of identifying integrable collective spin models using average EE in vanishing bipartitions is ruled out. 

\section{Average Entanglement Entropy in Lipkin-Meshkov-Glick model} \label{sec:average}

\begin{figure*}[t]
\includegraphics[width=\textwidth]{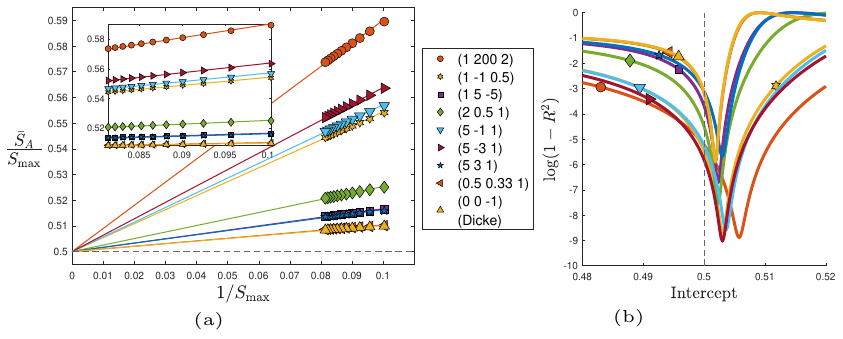}
\caption{(a) Normalized average entanglement $
            \frac{\bar S_A}{S_{\max}}$ at half bipartition $p=1/2$, as a function of the inverse maximal entanglement $1/S_{\max}=1/\log(N/2+1)$, given parameters $(\gamma_x \, \gamma_y \, h)$ in the LMG model. The linear fits correspond to $a+b/S_{\text{max}}$ with intercept $a$ fixed to be $1/2$. The inset shows a zoomed in version of the linear fit. (b) Plot of $\log (1-R^2)$ for all the parameter sets in (a), when the intercept $a$ is fixed at different values in $[0.48,0.52]$. Here, $R^2$ is the coefficient of determination of the linear fit for different intercept values. For $a=1/2$, $(1-R^2)$ varies between $10^{-3}$ and $10^{-6}$. The number of qubits $N \in [2\times10^3,10^4]$ for the data points in this figure, and the average EE is over the eigenstates from only the positive parity sector of $R_z^{\pi}$. Also, the eigenbasis of $H_{\text{LMG}}$ for $(\gamma_x \, \gamma_y \, h) = (0,0,-1)$ plotted here is the Dicke basis.} 
\label{fig:linfit}
\end{figure*}
We now introduce the LMG model, and then present our main result on its average EE: that the uniform normalized average of the half-bipartition EE of all eigenstates converges to a fixed value of $1/2$ in the thermodynamic limit, irrespective of the choice of parameters.

The Hamiltonian of a long-range (anisotropic) interacting system of $N$ spin-$1/2$ qubits in the presence of a transverse magnetic field of strength $h$ can be given by
\begin{align}\label{eq:HamiltonianLongRange}
    H=&-\frac{1}{4N} \sum_{\substack{k,l=1 \\ k\neq l}}^N (\frac{\gamma_x}{|k-l|^\alpha} \sigma_k^x \sigma_l^x + \frac{\gamma_y}{|k-l|^\alpha} \sigma_k^y \sigma_l^y) \nonumber\\&- \frac{h}{2} \sum_{k=1}^N \sigma^z_k.
\end{align}

For the case of infinite-range interaction, that is, $\alpha=0$, the system reduces to the well known LMG model \cite{ribeiro2007thermodynamical,santos2016excited,pappalardi2018scrambling}. The Hamiltonian can then be written as 
\begin{align}\label{eq:Hamiltonian}
    H_{\text{LMG}}=&-\frac{1}{N}(\gamma_x J_x^2+\gamma_y J_y^2)-h J_z,
\end{align}
in terms of the collective spin operators \eqref{SpinOperators} where $\gamma_x$ and $\gamma_y$ are real value parameters determining the respective interaction strengths. The square of the total angular momentum operator, $J^2$, commutes with  $H_{\text{LMG}}$. This implies that the eigenvalues of $J^2$, $j(j+1)$ and hence $j$, are constants of motion. The LMG model is integrable irrespective of the choice of the parameter set $\left(\gamma_x,\gamma_y,h \right)$.

The LMG Hamiltonian commutes with the rotation operator $R_z^{\pi} = \exp{\left(-i \pi J_z \right)}$ whose eigenvalues are $\pm1$. Thus the Hamiltonian can be block diagonalized into positive and negative parity sectors of $R_z^{\pi}$. All our numerical results for $H_{\text{LMG}}$ are in the corresponding positive parity sector .

We numerically study the average EE corresponding to the $N/2{:}N/2$ half bipartition ($p=1/2$) in the LMG model. We compute this average EE for several values of $N$ and parameter sets $\left(\gamma_x,\gamma_y,h \right)$. For the special case of $\gamma_x{=}\gamma_y{=}\gamma$,  $H_{\text{LMG}} = -\frac{\gamma}{N}J^2 +  \frac{\gamma}{N}J_z^2 -hJ_z$, which commutes with both $J^2$ and $J_z$. Hence the Dicke basis is the eigenbasis of $H_{\text{LMG}}$ for $\gamma_x = \gamma_y = \gamma$. This eigenbasis is nondegenerate for $h\neq0$. This, together with Eq. \eqref{eq:entone} implies that for $\gamma_x = \gamma_y = \gamma$ and $h\neq0$, the normalized average EE of the eigenbasis of $H_{\text{LMG}}$ 
converges to $1/2$ for $p>0$. Moreover, the equal superposition Dicke basis (introduced in Sec. \ref{sec:Dicke}) is the eigenbasis of the LMG model (up to a global rotation) for the following choices of parameter sets $(\gamma_x,\gamma_y,h) = (\gamma,0,0)$, $(0,\gamma,0)$, and $(\gamma,\gamma,0)$ where $\gamma$ is a real nonzero number. Hence the EE analysis of the equal superposition Dicke basis in Sec. \ref{sec:Dicke} applies to the LMG model with these parameter sets.

For the more general case of $\gamma_x\neq \gamma_y$ and $h\neq 0$, we compute $\bar{S}_A$ numerically for half-bipartition, and analyze its thermodynamic limit using finite-size scaling for various choices of parameters $\left(\gamma_x,\gamma_y,h \right)$. The results are shown in Fig. \ref{fig:linfit}(a), where we see that the normalized average EE decreases with increasing system size, and that finite-size scaling shows that it always approaches a thermodynamic limit value of $1/2$. This corresponds to the ``Dicke" value $\overline{S}_A/S_{\max} \rightarrow \frac{1}{2}$.
Moreover, the linear scaling in Fig. \ref{fig:linfit}(a) implies that
\begin{equation}
    \frac{\overline{S}_A}{S_{\max}} \simeq  \frac{1}{2} +\mathcal{O}\left(\frac{1}{S_{\max}} \right),
\end{equation}
consistent with the sub-leading term in Eq. \eqref{eq:entone}.  This is confirmed by an analysis of the coefficient of determination measuring the quality of the fit, shown in Fig. \ref{fig:linfit}(b) (similar to the analysis of \cite{leblond2019entanglement}). We see that the best linear fits occur when the intercept (that is, the value in the thermodynamic limit) is fixed near $1/2$. The results shown in the figure are for number of qubits $N$ up to $10^4$.

In Fig. \ref{fig:linfit}(a), the data from $\left(\gamma_x,\gamma_y,h \right) = \left(0,0,-1\right)$ corresponds to the Dicke basis (in the positive $R_y^{\pi}$ sector) for $N\in [2000,10000]$. We observe that the finite-size scaling behaviour of the Dicke basis EE is similar to that of other parameters of the LMG model.  Also, the intercept that gives the best fit, as measured by the coefficient of determination (see the minima in Fig. \ref{fig:linfit}(b)), deviates from $1/2$ due to finite-size effects. The deviation shown is very similar to that of the LMG model. This evidence further supports our conclusion that the normalized average EE for the LMG is the same as the Dicke basis in the thermodynamic limit. 

The results shown here are all for eigenstates within the symmetric subspace. Since $H_{\text{LMG}}$ commutes with $J^2_{total}$, the dynamics in different total spin $j$ sectors ($j$ ranging from 0 to $\frac{N}{2}$ for even $N$) will be decoupled to each other. Thus, in principle, it is possible to calculate average EE in different total spin $j$ sectors. However, numerical computations in other sectors will be much more computationally costly.

\begin{figure}
    \centering
    \includegraphics[width=0.48\textwidth]{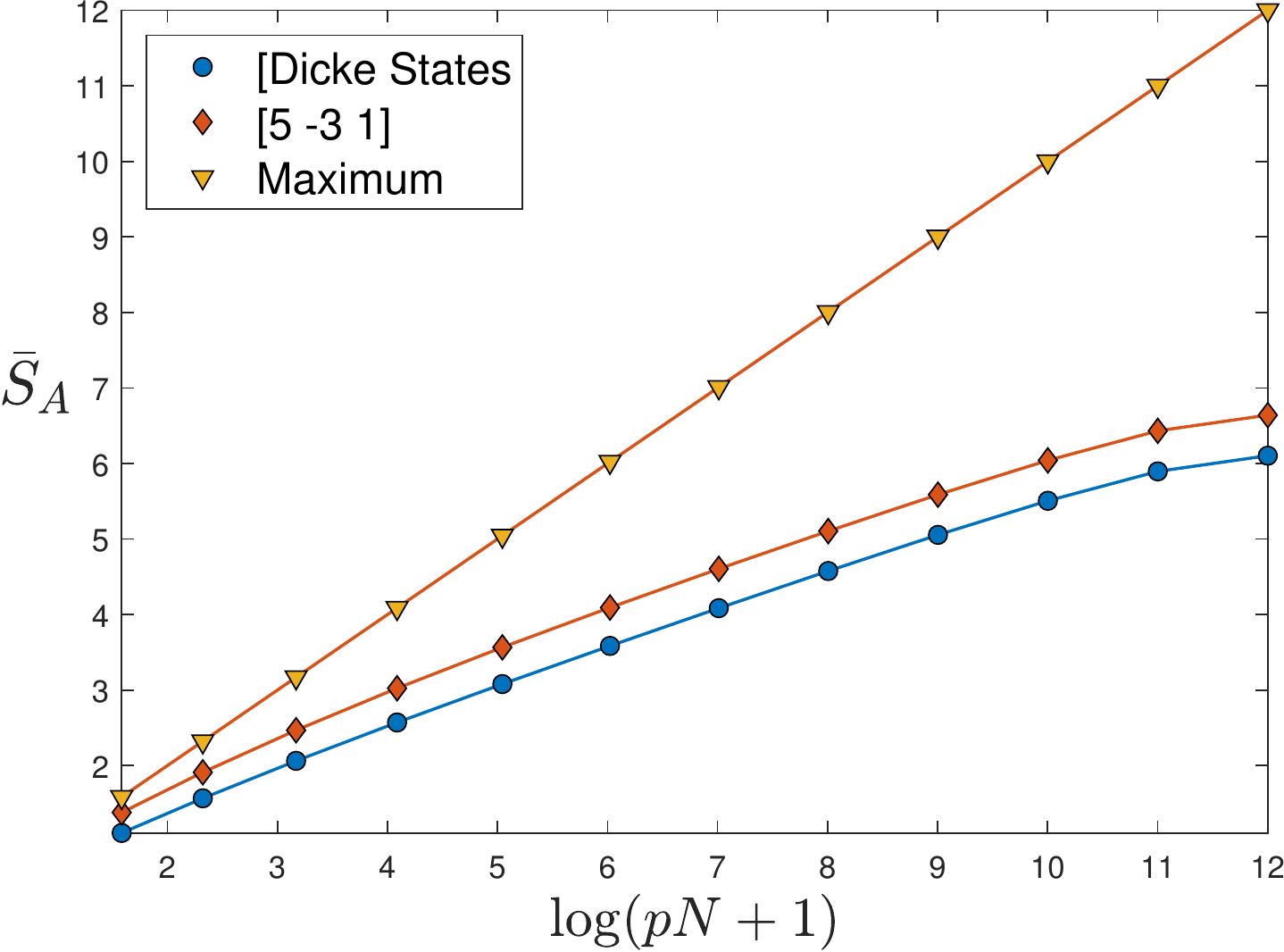}
    \caption{Plot of the average EE for different bipartitions $p=N_A/N$, of the eigenstates in LMG model corresponding to $\left(\gamma_x,\gamma_y,h \right) = \left(5,-3,1\right)$ and for the Dicke basis. The maximum EE in the given bipartition is also plotted for comparison. $N=8192$ $(\equiv 2^{13})$ for this plot. Up to close to the half bipartition $p=1/2$, all the curves are roughly straight lines.}
    \label{fig:subs}
\end{figure}

We also carried out the numerical analysis for another bipartition $p=1/4$, for which we have observed that $\frac{\overline{S}_A}{S_{\max}} \rightarrow \frac{1}{2}$. This, together with the Dicke basis result in Eq. \eqref{eq:entone}, strongly suggest that for the LMG the ratio $\frac{\overline{S}_A}{S_{\max}}$ is fixed for any non-vanishing bipartition. We confirm this with our results in Fig. \ref{fig:subs}, where we plot the average EE as a function of the subsystem size $p$ for $N=2^{13}$, and find an almost linear growth consistent with a fixed ratio. Mathematically, this implies that $c_0(p)$ is a fixed value independent of the choice of $p>0$, where $c_0(p)$ is defined by
\begin{eqnarray} \label{eq:VolLawCoeff}
\overline{S}_A(p) = c_0(p) S_{\text{max}}(p), 
\end{eqnarray}
where $S_{\text{max}}(p)=\log(pN+1)$. This establishes an interesting difference with other integrable lattice systems where it has been found that $c_0(p)$, that is, the coefficient of the leading term in the volume law of entanglement, does depend on $p$ \cite{hackl2019average,lydzba2101entanglement,lydzba2020eigenstate,leblond2019entanglement} (unlike in chaotic systems \cite{vidmar2017entanglement2}). 
In fact, that dependence can be calculated exactly for random quadratic systems \cite{lydzba2020eigenstate}. 

\section{Entanglement distribution in the spectrum}\label{sec:distribution}

\begin{figure}
    \centering
    \includegraphics[width=0.5\textwidth]{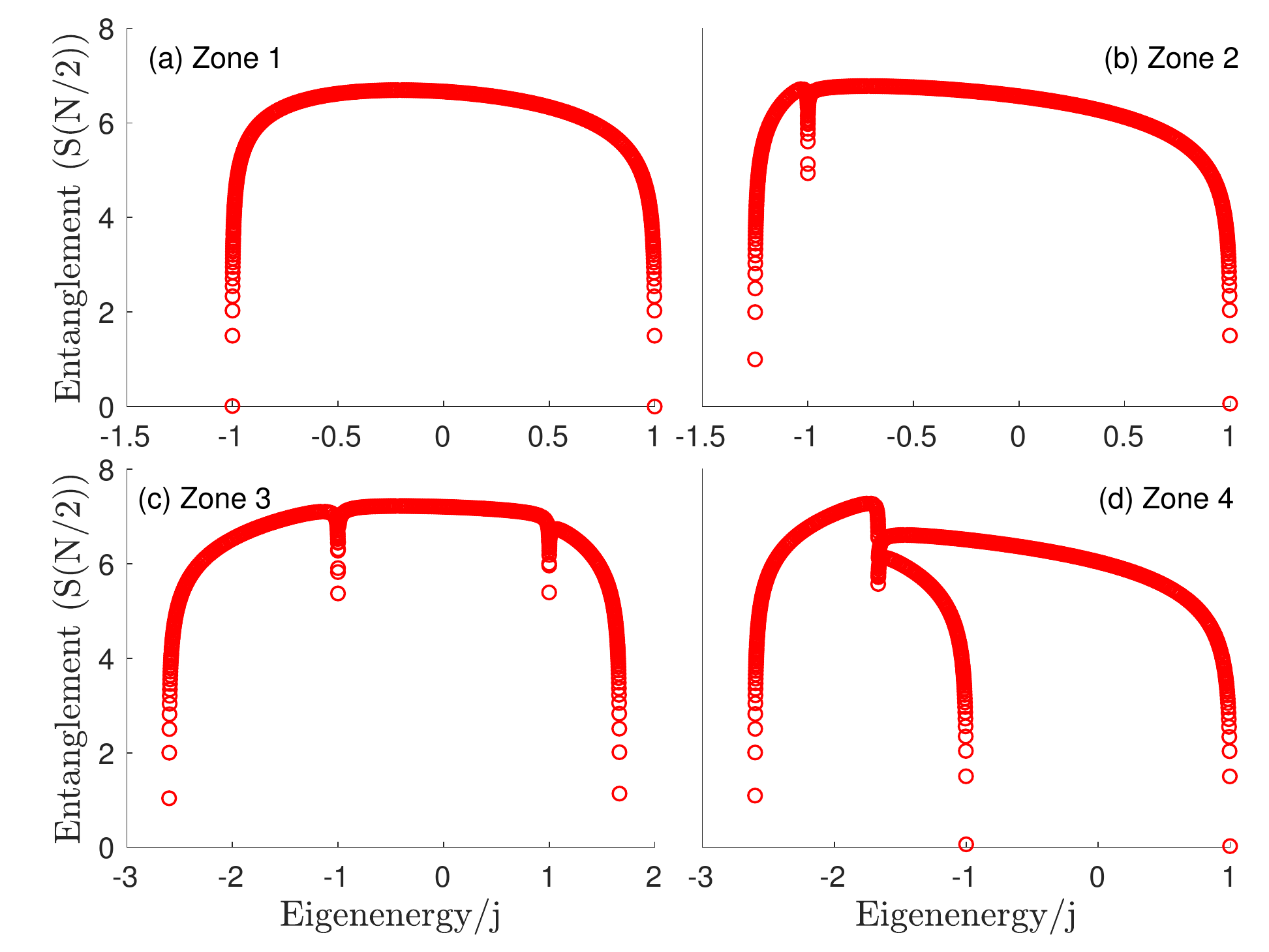}
    \caption{Distribution of entanglement entropy at half-bipartition for all the eigenstates of the LMG model in the positive parity sector, as a function of their eigenenergy. Zones $1$ to $4$ correspond to the choice of four sets of parameters $(\gamma_x,\gamma_y,h) = (1/2,1/3,1), (2,1/2,1), (5,-3,1),$ and $(5,3,1)$ in the model \cite{ribeiro2007thermodynamical,ribeiro2008exact}. There are dips in the entanglement distribution at eigenenergies corresponding to the singularities in the density of states. For this plot, we have $N=10^4$.}
    \label{fig:EEzones}
\end{figure}

In order to further understand the universality of the average EE shown in Sec. \ref{sec:average}, we study the entanglement distribution in the spectrum, plotted in Fig. \ref{fig:EEzones}. We also explore the properties of the LMG model in order to better understand the features of this distribution. 

The density of states (DOS) in the LMG model exhibits singularities (logarithmic divergences) and discontinuities \cite{ribeiro2007thermodynamical, ribeiro2008exact}. These are known to be caused by the so-called Excited-state quantum phase transitions (ESQPTs) \cite{cejnar2020excited}. These ESQPTs have been seen in many models with finite degrees of freedom, and are most often associated with unstable fixed points of the corresponding classical system \cite{cejnar2020excited,santos2016excited}.

To better understand ESQPTs in the LMG model, and their effect on the entanglement distribution, we study the classical LMG in Appendix \ref{app:classical}. We first derive the classical equations of motion. Then, we find a list of fixed points and analyze their stability. Based on the existence and stability of these fixed points (or equivalently, on the qualitative behavior of DOS), the parameter space can be divided into 4 different ``zones'' as described in Appendix \ref{app:classical} \cite{ribeiro2007thermodynamical, ribeiro2008exact,nader2021avoided}. The plots in Fig. \ref{fig:EEzones} correspond to each of these zones.

There is an ESQPT in the quantum LMG model corresponding to every unstable fixed point in the classical LMG model. For finite-$j$ values, the energy at which these ESQPTs occur in the quantum case are very close to the classical Hamiltonian value at these unstable fixed points. In Fig. \ref{fig:EEzones}, we study half-bipartition EE as a function of the eigenenergy for different parameter sets in all the four zones. We observe that the behavior of EE has significant differences across different zones. Specifically, the entanglement in the eigenstates close to the ESQPT energy deviates from the entanglement in the bulk of the spectrum in all the four zones. These small deviations are in addition to the more significant deviations in the EE at the edges of the spectrum. 

Nonetheless, ESQPTs do not seem to significantly affect the EE in the bulk of the spectrum. The thermodynamic limit of the normalized average EE converges to the same value for parameter sets across different zones, likely due to the much higher number of eigenstates in the bulk. This is evident from Fig. \ref{fig:linfit} where we have shown results with parameter choices in all the four zones. Specifically, $(\gamma_x,\gamma_y,h) = (1/2,1/3,1), (2,1/2,1), (5,-3,1),$ and $(5,3,1)$ correspond to zones 1,2,3, and 4, respectively, as studied in \cite{ribeiro2008exact}. In all cases, the convergence to the value of $1/2$ is similar, from which we deduce that the thermodynamic limit of the normalized average EE is independent of the DOS zone.

\section{Conclusion}\label{sec:conclusion}

We have studied the entanglement entropy in the eigenstates of a collective spin \emph{integrable} model, that is, the LMG model. We have numerically shown that the average EE in non-vanishing bipartitions converges to a value in the thermodynamic limit which corresponds to that of the Dicke basis. We have also analytically calculated this Dicke value to be half of the maximal possible EE. This shows that the LMG eigenstates have EE that is far from maximal, in contrast to random permutation-symmetric states for which $\bar S_A/ S_{\max} \simeq 1$ \cite{stockton2003characterizing,seshadri2018tripartite}. This behavior of the average EE in the LMG model is dramatically different to that of \emph{chaotic} collective spin models, such as the quantum kicked top \cite{kumari2021entanglement}. A similar picture has been observed in the context of integrable versus chaotic models in one dimension. This suggests that this quantity being far away from the maximal could serve as a good indicator of the absence of quantum chaos in many-body quantum systems.

The system studied here belongs to one of the various classes of solvable/integrable models known in the literature. This includes free fermions, interacting models in 1D such as (nearest-neighbor) XXZ and Fermi-Hubbard, and long-range interacting models such as Haldane-Shastry and Calogero-Sutherland models. Each of these classes is very different from the others, both in their physical features and in the method to solve them. Our results, together with previous ones for 1D systems \cite{vidmar2017entanglement,hackl2019average,lydzba2020eigenstate,leblond2019entanglement}, suggest that integrable systems have an average entanglement value that is far from maximal (with a normalized average value much smaller than 1) and this characteristic could be used to distinguish them from chaotic systems. Nonetheless, different classes of integrable models can have different values of average EE in the thermodynamic limit while still being far away from the maximal. This also opens up the possibility that the exact value of the average EE characterizes the class of integrable models. Beyond the results of this paper, the best evidence we currently have for this is the analysis in \cite{leblond2019entanglement}, which shows that different 1D models have a very similar average EE. It would be interesting to explore this idea further via analytical and numerical calculations in different known integrable models.

This potential measure of integrability in terms of average EE in the eigenstates has some other advantages over other widely studied indicators. Firstly, it does not require any initial quantum state for its study. This contrasts with the entanglement growth in integrable versus chaotic quantum systems, whose analysis depends on the choice of initial quantum state in the system \cite{lombardi2011entanglement,ruebeck2017entanglement,chaudhury2009quantum,Neill_2016}. Secondly, integrable systems may have a few instabilities insufficient to render the system chaotic, such as those that lead to ESQPTs in the LMG model \cite{ribeiro2007thermodynamical,ribeiro2008exact}. The average EE value in the thermodynamic limit does not seem to be affected by them in integrable systems, as we observed in the present model. This is an advantage over measures that are significantly affected by those few instabilities, for example, out-of-time-ordered correlators (OTOCs) that have been shown to exhibit chaos-like behavior in the LMG model despite it being integrable \cite{Hirsch2020}.

We have also computed the distribution of entanglement across the energy spectrum for various system parameters. Despite the average being constant, we found that the distributions display different singular points that correspond to the ESQPTs. This could stimulate further work connecting entanglement with ESQPTs, in a way similar to what happens with ground state phase transitions \cite{Vidal_2003,latorre2005entanglement}. 

The LMG model, as well as other collective spin models, are currently the subject of quantum simulation studies \cite{Rey2019,munoz2020simulating,Sieberer_2019}, in particular, with optical lattices \cite{Martin_2013}. It is expected that their integrability will have an impact on the efficiency of their simulation \cite{Sieberer_2019}. We hope that the present results help us in understanding the physical features that have direct consequences on the efficiency of quantum simulators.

\begin{acknowledgments}
The authors would like to thank Marcos Rigol for insightful discussions. MK would also like to thank V. Ravi Chandra, Cheng-Ju Lin, Jack Davis and Namit Anand for useful discussions. AMA acknowledges funding from the Alexander von Humboldt Foundation. This research was supported in part by Perimeter Institute for Theoretical Physics. Research at Perimeter Institute is supported in part by the Government of Canada through the Department of Innovation, Science and Economic Development and by the Province of Ontario through the Ministry of Colleges and Universities. 
\end{acknowledgments}

\bibliography{references}
\bibliographystyle{quantum}

\widetext
\appendix


\section{Entanglement of Dicke states and the LMG model} \label{app:dicke}

Here, we prove Eq. \eqref{eq:entupper}, \eqref{eq:entlower} and \eqref{eq:entone} in the main text. In order to do this, we need to estimate the entanglement entropy of the Dicke states $\ket{j,m}$. For a bipartition in which w.l.o.g the smallest subsystem size is $N_A$ they have Schmidt coefficients $\{\lambda_q \}$ which can be written in terms of the \emph{hypergeometric distribution} as \cite{latorre2005entanglement,moreno2018all}
\begin{equation}
\lambda_q= \frac{{{N_A}\choose{q}} {{ N_B}\choose{j-m-q}}} {  {{2j}\choose{j-m}}} \quad q \in{0,1,...,j-m}.
\label{SchmidtCoeffDicke2}
\end{equation}

\subsection{Regions proportional to system size}\label{app:dickehalf}

For $N_A \propto N=2j$ the entropy of the hypergeometric distribution (and the EE of Dicke states) is approximated by \cite{cichon2013delta}
\begin{equation}\label{eq:approx}
S(\text{Tr}_B[\ket{j,m}\bra{j,m}])\equiv S(N, p_1 N, p_2 N ) =\frac{1}{2}\log(2 \pi e N p^*_1 p^*_2) +\frac{
\log e }{12N}\left( -10 +\frac{4}{p^*_1}+\frac{4}{p^*_2}-\frac{1}{p^*_1 p^*_2} \right)+ \mathcal{O}\left(\frac{1}{N^2}\right),
\end{equation}
where $p_i^*=p_i(1-p_i)$, $p_1=(j-m)/N$ and $p_2= N_A/N$ and the logs are in base $2$. This approximation is useful when both $p_1,p_2$ are $o(N)$, so that the second term is not too large. 

Let us now calculate upper and lower bounds to the average entanglement entropy, averaged over all $m$. First, notice that the largest value at any $p_2$ occurs at $p_1=1/2$, so that
\begin{equation}
S(N,p_1 N,p_2 N) \le S(N,N/2,p_2 N)=\frac{1}{2}\log(\frac{\pi e N p_2^*}{2})-\frac{\log e}{2N}+\mathcal{O}\left( \frac{1}{N^2}\right).
\end{equation}
This is thus an upper bound for the average
\begin{equation}\label{eq:updicke}
\sum_m \frac{1}{N+1}S(N,j-m,p_2 N)\le \frac{1}{2}\log(\frac{\pi e N p_2^*}{2})-\frac{\log e}{2N}+\mathcal{O}\left( \frac{1}{N^2}\right).
\end{equation}
Substituting $N=2j$ proves Eq. \eqref{eq:entupper}. We now compute the lower bound by eliminating a number of positive terms from the average, such that the expression above can be used without too large an error. Let us write

\begin{align}\label{eq:lowb}
\sum_{m=-j}^{j} \frac{1}{N+1}&S(N,j-m,p_2 N)\ge \frac{1}{N+1}\sum_{j-m=\epsilon N/2}^{N(1-\epsilon/2)}S(N,j-m,p_2 N) \\ &\ge \frac{1}{2(N+1)} \left(\sum_{k=\epsilon N/2}^{N(1-\epsilon/2)} \log (2 \pi e  p_2^* k (1-\frac{k}{N}))\right) -\frac{\log e}{12(N+1)}\left(10 +\frac{1}{\epsilon p^*_2} \right)+ \mathcal{O}\left(\frac{1}{N^2}\right) .
\end{align}
In the last line we have replaced $k=j-m$ for simplicity.
 We now calculate the sum over $k$ using the Euler-McLaurin formula, which states that \begin{align}\label{eq:emc}
\sum_{k=l_0}^{l_1} f(k)-\int_{l_0}^{l_1} f(x) \text{d}x=\frac{1}{2}(f(l_0)+f(l_1))+\frac{1}{12}(f'(l_1)-f'(l_0))+\rho(f;,l_0,l_1),
\end{align}
where $\vert \rho(f;l_0,l_1)\vert \le \frac{1}{120}\int_{l_0}^{l_1} \vert f'''(x)\vert \text{d}x $. Using Eq. \eqref{eq:emc}, we can identify
\begin{align}
&f(x)=\log (\frac{x}{N} (1-\frac{x}{N})), \,\, l_0=\frac{N \epsilon}{2}, \,\,\, l_1=N(1-\frac{\epsilon}{2}) \\
&f(N\epsilon/2)=f(N(1-\epsilon/2))=\log \frac{\epsilon}{2}(1-\frac{\epsilon}{2}) \\
& f'(N(1-\epsilon/2))-f'(N \epsilon/2)=\frac{1}{\log e}\left( \frac{2}{N(1-\epsilon/2)}-\frac{2}{N \epsilon} \right)\\
&\int_{l_0}^{l_1} \vert f'''(x)\vert \text{d}x \le  \mathcal{O}\left( \frac{1}{N \epsilon} \right).
\end{align}
The leading order correction is given by the term $\frac{1}{2}(f(N \epsilon/2)+f(N(1-\frac{\epsilon}{2}))$, so we can write
\begin{equation}
\sum_{k=\epsilon N/2}^{N(1-\epsilon/2)} \log[\frac{k}{N}(1-\frac{k}{N})] =N \int_{\epsilon/2}^{1-\epsilon/2} \log[y(1-y)] \text{d}y+\mathcal{O}(\log \epsilon).
\end{equation}
 After calculating the integral we obtain
\begin{align}
\sum_{k=\epsilon N/2}^{N(1-\epsilon/2)} \log[\frac{k}{N}(1-\frac{k}{N})] &=N\left(4(\epsilon-1) \log 4e^2 +(2-\epsilon)\log (2-\epsilon)-\epsilon \log \epsilon \right) + \mathcal{O}(\log \epsilon) \\&\ge -N(1-\epsilon)\log 4e^2-N \epsilon \log \epsilon +\mathcal{O}(\log \epsilon).
\end{align}
Thus we get that 
\begin{align}
\sum_{k=\epsilon N/2}^{N(1-\epsilon/2)} \log (2 \pi e  p_2^* k (1-\frac{k}{N}) &\ge N (1-\epsilon) \log(2\pi e p_2^* N) -N(1-\epsilon)\log 4e^2-N \epsilon \log \epsilon +\mathcal{O}(\log \epsilon) \\ & \ge N(1-\epsilon) \left(\log \frac{\pi N p_2^*}{2e} \right) - N \epsilon \log {\epsilon}+\mathcal{O}(\log \epsilon).
\end{align}
Now, let us choose an $\epsilon \propto N^{-1/2}$. In that case, the last term above is
\begin{equation}
N \left(\log \frac{\pi N p_2^*}{2e} \right) +\mathcal{O}(N^{1/2} \log N) .\end{equation}
Putting everything together, and accounting for all the errors, we obtain the lower bound Eq. \eqref{eq:entlower}
\begin{align}
\sum_{k=0}^{N} \frac{1}{N+1}S(N,k,p_2 N)&\ge \frac{N}{2(N+1)}\log \frac{\pi N p_2^*}{2e} +\mathcal{O}( (p_2^*)^{-1}  N^{-1/2})+\mathcal{O}( \log (N)  N^{-1/2}).
\end{align}
We see that in the limit of large $N$ it converges to the same value as the upper bound in Eq. \eqref{eq:updicke}. Thus in this thermodynamic limit, given that $S_{\max}=\log(p_2N+1)$, it is easy to see that
\begin{equation}\label{eq:limitaverage}
\lim_{N \rightarrow \infty} \frac{1}{(N+1) }\sum_{k=0}^{N} \frac{S(N,k,p_2 N)}{S_{\text{max}}}=\frac{1}{2},
\end{equation}
which is independent of $p_2$ (defined as $p$ in the main text).

\subsection{Single qubit subsystem} \label{app:1QubitDicke}

\subsubsection*{Dicke basis}

For a single qubit bipartition, that is, $N_A{:}N = 1{:}(N-1)$, we have a simple closed expression of the entanglement of any Dicke state $\ket{N,k}$ \cite{moreno2018all}
\begin{equation}
S(\text{Tr}_B[\ket{N,k}\bra{N,k}])\equiv S(N,k,1)=-\frac{k}{N}\log \left(\frac{k}{N}\right)-\left(1-\frac{k}{N}\right)\log \left(1-\frac{k}{N}\right).
\end{equation}
We now want to calculate the average over all k
\begin{equation}
\sum_{k=0}^N \frac{S(N,k,1)}{N+1}.
\end{equation}
Let us write $f(k)=S(N,k,1)$. We use again the Euler-McLaurin formula from Eq. \eqref{eq:emc}. First, notice that $f(x)$ is not differentiable at $x=\{0,n\}$, so we will instead choose $l_0=1,l_1=N-1$, since $\sum_{k=0}^{N} f(k)=\sum_{k=1}^{N-1} f(k)$. We then have the following expressions
\begin{align}
&f(x)=-\frac{x}{N}\log \left(\frac{x}{N}\right)-\left(1-\frac{x}{N}\right)\log \left(1-\frac{x}{N}\right) \\ &
f'(x)=\frac{1}{N}\left(\log(1-x/N)-\log(x/N) \right)
\\ & f'''(x)=\left(\frac{1}{N x^2}-\frac{1}{N^3 \left(1-\frac{x}{N}\right)^2} \right) \log e.
\end{align}
We can calculate every term of Eq. \eqref{eq:emc} straightforwardly as
\begin{align}
&\int_{l_0}^{l_1} f(x) \text{d}x= \left(\frac{N}{2}-1 \right)\left( 2 \log N +\log e \right)-\frac{(N-1)^2}{N}\log(N-1) \\
&f(N-1)=f(1)=-\frac{1}{N}\log \left(\frac{1}{N}\right)-\left(1-\frac{1}{N}\right)\log \left(1-\frac{1}{N}\right) \\
&f'(N-1)-f'(1)=\frac{2}{N}\left(\log(\frac{1}{N})-\log(1-\frac{1}{N}) \right)\\
& \int_{1}^{N-1}\vert f'''(x) \vert \text{d}x = \frac{2(N-2)^2}{N^2(N-1)} \log e \le \frac{2}{N} \log e.
\end{align}
This way we see that as $N$ large enough this approximates
\begin{equation}
\frac{1}{N+1}\sum_{k=0}^{N} S(N,k,1)=\frac{\log e}{2}+ \mathcal{O}\left(\frac{\log N}{N}\right),
\end{equation}
so in the limit, since $S_{\text{max}}=1$,
\begin{equation}
\lim_{N \rightarrow \infty} \frac{1}{(N+1) S_{\text{max}}}\sum_{k=0}^{N} S(N,k,1)=\frac{\log e}{2S_{\text{max}}}=\frac{\log e}{2} \simeq 0.72.
\end{equation}

\subsubsection*{Equal superposition Dicke basis}
For any $N-$qubit permutation symmetric pure state $\ket{\psi}$, the 1-qubit reduced state, $\varrho_A$ is given by \cite{wang2002pairwise, Kumari2017}
\begin{equation}
\varrho_A = \begin{bmatrix} v_++w & x_+^*+x_-^* \\ x_++x_- & v_-+w \end{bmatrix}
\end{equation}
where
\begin{eqnarray}\label{eq:Elements}
v_{\pm} = \frac{N^2-2N+4 \langle J_z^2 \rangle \pm 4 \langle J_z \rangle (N-1)}{4N(N-1)}, 
x_{\pm} = \frac{(N-1)\langle J_+ \rangle \pm \langle [J_+,J_z]_+ \rangle}{2N(N-1)}, \text{ and }
w =\frac{N^2-4\langle J_z^2 \rangle}{4N(N-1)}.
\end{eqnarray}
The two eigenvalues of $\varrho_A$ are calculated to be $\lambda_{\pm} = \frac{1}{2}\pm \frac{\sqrt{ \langle J_z \rangle^2 + |\langle J_+ \rangle|^2}}{N}$ using \eqref{eq:Elements}. For any state in the equal superposition Dicke basis, $\{ \{ \frac{1}{\sqrt{2}} (|j,m\rangle \pm |j,-m\rangle) \}_{m=1}^j, \ket{j,0} \}$, $\langle J_z \rangle = 0$ and $\langle J_+ \rangle = 0$. This implies that both the eigenvalues of $\varrho_A$ are $1/2$, making the 1-qubit EE of every state in this basis equal to the maximal value of 1. 


\section{Classical analysis of LMG model}
\label{app:classical}
Here, we derive the classical equations of motion for the LMG Hamiltonian \eqref{eq:Hamiltonian}, and study the fixed points of the classical LMG model. The angular momentum vector, $\vec{J}$, is a cross product of position and momentum vectors, that is, $\vec{J} = \vec{r} \times \vec{p}$. In terms of the components of the vectors, $J_x = yp_z - zp_y$, $J_y = zp_x - xp_z$, and $J_z = xp_y - yp_x$. Thus, the LMG Hamiltonian \eqref{eq:Hamiltonian} can be re-expressed as \begin{equation}
    H = - \frac{1}{N} \left(\gamma_x (yp_z - zp_y)^2 +\gamma_y (zp_x - xp_z)^2 \right) -h(xp_y - yp_x).
    \label{LMG2}
\end{equation}
Using the Hamilton's equation of motion, $\dot{q_i} = \frac{\partial H}{\partial p_i}$ and $\dot{p_i} = -\frac{\partial H}{\partial q_i}$, we obtain

\begin{subequations}
\begin{align*}
\dot{x} = & - \frac{2}{N}\gamma_y zJ_y + hy,   & 
\dot{p}_x = & - \frac{2}{N}\gamma_y p_zJ_y + hp_y, \\
\dot{y} =  & \frac{2}{N}\gamma_x zJ_x - hx, &
\dot{p}_y = & \frac{2}{N}\gamma_x p_zJ_x - hp_x, \\
\dot{z} = & - \frac{2}{N}\left( \gamma_x yJ_x - \gamma_y xJ_y \right),  &
\dot{p}_z = & -\frac{2}{N}\left(\gamma_x p_yJ_x - \gamma_y p_xJ_y \right).  
\end{align*}
\end{subequations}
Using these equations, the classical equations of motion for the components of the angular momentum vector can be derived as
\begin{subequations}
\begin{align*}
\dot{J_x} & = J_y \left( h - \frac{2}{N}\gamma_y J_z\right), &
 \dot{J_y} & = J_x \left( \frac{2}{N}\gamma_x J_z - h\right), &
 \dot{J_x} & = \frac{2}{N}J_xJ_y \left(\gamma_y-\gamma_x\right). 
\end{align*}
\end{subequations}

Furthermore, $\frac{d|\vec{J}|^2}{dt} = \vec{J}\cdot \frac{d\vec{J}}{dt} = 0$ implying that $|\vec{J}| = j$ is a constant of motion for the LMG model and the classical evolution is on a sphere of radius $j$. Motivated from the quantum study of the LMG model, we substitute the constant $N=2j$ in the Hamiltonian \eqref{eq:Hamiltonian}. Additionally, substituting $X=J_x/j$, $Y=J_y/j$ and $Z=J_z/j$, we get the rescaled equations of motions to be
\begin{align}
\dot{X} & = Y \left(h-\gamma_yZ\right), &
 \dot{Y} & = X \left( \gamma_xZ-h \right), & 
 \dot{Z} & = XY(\gamma_y-\gamma_x). \label{ClassicalLMGeqs} 
\end{align}

  \begin{table}[t]
        \begin{tabular}{ccl|c|c}
        &&& $H_0$ & Existence conditions (if any) \\
        \hline
        $\text{FP}_{\text{XZ}}^{\pm}$ & = & $\left(\pm \sqrt{1-\left( \frac{h}{\gamma_x}\right)^2},0,\frac{h}{\gamma_x} \right)$ & $-\frac{h^2+\gamma_x^2}{2\gamma_x}$ & $|h| < |\gamma_x|$\\
        $\text{FP}_{\text{YZ}}^{\pm}$ & = & $\left(0, \pm \sqrt{1-\left( \frac{h}{\gamma_y}\right)^2},\frac{h}{\gamma_y} \right)$ & $-\frac{h^2+\gamma_y^2}{2\gamma_y}$ & $|h| < |\gamma_y|$  \\
        $\text{FP}_{\text{Z}}^{\pm}$ & = & $(0,0,\pm 1)$ & $\mp 1$ 
    \end{tabular}
    \caption{Fixed points and periodic orbits of the classical LMG model [in the form $(X,Y,Z)$]. $H_0$ denotes the value of the classical LMG Hamiltonian at the corresponding FP $(X,Y,Z)$.}
    \label{FPlist1}
  \end{table}
  
We obtain 6 fixed points (FPs) listed in Table \ref{FPlist1} using $\dot{X}=0$, $\dot{Y}=0$, and $\dot{Z}=0$. $\text{FP}_{\text{XZ}}^{\pm}$ and $\text{FP}_{\text{YZ}}^{\pm}$, are degenerate, respectively. The existence and stability of these FPs vary across the parameter space $(\gamma_x,\gamma_y,h)$. Based on the behavior of the DOS, the parameter space $(\gamma_x,\gamma_y,h)$ can be divided into 4 ``zones" \cite{ribeiro2007thermodynamical, ribeiro2008exact,nader2021avoided}, each with a qualitatively different DOS. Here, we follow the numbering of the zones of \cite{ribeiro2008exact}. Without loss of generality, we assume $h>0$ in the following.
\begin{itemize}
    \item Zone I: $\vert \gamma_x \vert < h$ and $\vert \gamma_y \vert < h$. Only the Z-pole FPs $\text{FP}_Z^{\pm}$ exist and both these FPs are stable. Hence, there is no divergence in DOS. 
    \item Zone II: (a) $\vert \gamma_x \vert < h < \vert \gamma_y \vert $, (b) $\vert \gamma_y \vert < h < \vert \gamma_x \vert $. One of $\text{FP}_Z^{\pm}$ is stable and the other is unstable. There is a divergence in the DOS at $H_0$ value corresponding to the unstable z-pole FP. Additionally, $\text{FP}_{\text{YZ}}^{\pm}$ and $\text{FP}_{\text{XZ}}^{\pm}$ exist in (a) and (b), respectively, and are stable. 
    \item Zone III: (a) $h < -\gamma_y $ and $h < \gamma_x$, (b) $h < -\gamma_x <$ and $h < \gamma_y$. Both Z-pole FPs $\text{FP}_{\text{Z}}^{\pm}$ are unstable fixed points. There are two divergences in the DOS at $H_0$ values corresponding to these FPs, that is, $H_0 = \pm1$. Additionally, $\text{FP}_{\text{XZ}}^{\pm}$ as well as $\text{FP}_{\text{YZ}}^{\pm}$ exist in this zone and are stable. 
    \item Zone IV: (a) $h < \gamma_x $ and $h < \gamma_y$, (b) $h < -\gamma_x $ and $h < -\gamma_y$. Both the Z-pole FPs $\text{FP}_Z^{\pm}$ are stable. Both $\text{FP}_{\text{XZ}}^{\pm}$ and $\text{FP}_{\text{YZ}}^{\pm}$ exist in this zone, and one of these two pairs are stable and the other is unstable. There is a divergence in the DOS at $H_0$ value corresponding to the unstable pair.
\end{itemize}
The qualitative behavior of the entanglement distribution across the spectrum (Fig. \ref{fig:EEzones} in the main text) can be inferred from this classical analysis of the model, and in particular, from the knowledge of fixed points, their stability and their degeneracy.
\end{document}